\documentclass[10pt,a4paper]{article}
\usepackage[margin=2cm]{geometry}
\usepackage{cite}
\usepackage{amsmath,amssymb,amsfonts}
\usepackage{algorithmic}
\usepackage{graphicx}
\usepackage{textcomp}
\usepackage{xcolor}
\def\BibTeX{{\rm B\kern-.05em{\sc i\kern-.025em b}\kern-.08em
    T\kern-.1667em\lower.7ex\hbox{E}\kern-.125emX}}
\begin{document}

\title{Influence of dispersive element on phase noise suppression in Talbot effect based optical upconversion scheme
\thanks{This work has been funded by German Research Foundation (DFG) in the framework of priority program SPP2111 under project number 403187113 (Tabasco). This is the author's version (with permission of the organizers) that has been published in Proc. 21th ITG-Symposium on Photonic Networks, ISBN 978-3-8007-5423-6. The final version of record is available at https://www.vde-verlag.de/buecher/455423/itg-fb-294-photonische-netze.html}}

\author{Niels Neumann\\
\textit{Chair for RF and Photonics} \\
\textit{Engineering, TU Dresden}\\
01062 Dresden, Germany\\
niels.neumann@tu-dresden.de
\and
Zaid Al-Husseini\\
\textit{Chair for RF and Photonics} \\
\textit{Engineering, TU Dresden}\\
01062 Dresden, Germany
\and
Dirk Plettemeier\\
\textit{Chair for RF and Photonics} \\
\textit{Engineering, TU Dresden}\\
01062 Dresden, Germany}

\maketitle

\begin{abstract}
The temporal Talbot effect supports the generation of RF signals with high purity in optical domain. This allows for example a remote RF generation without the need for costly high-end electronic circuits and resource sharing. However, one of the most interesting features of the approach is its inherent ability to suppress phase noise during the carrier generation process. Besides the comb laser source, the dispersive element is the key component in this upconversion scheme. Ideally, it provides the right amount of dispersion over the whole spectral range of the comb. Broader combs are expected to allow a higher degree of phase noise suppression. The simulation of phase noise of an RF tone generated by an optical scheme is challenging in terms of computational and memory effort. Therefore, a simulation tool taking advantage of the properties of the comb and the Talbot effect had to be developed. In this paper, the dependency of the phase noise suppression on the dispersion characteristic is investigated yielding an important input to the design of these elements. First, the implemented simulation tool and the simulation parameters ensuring a correct estimation of the phase noise before and after dispersion influence are introduced. Then, the effect of different dispersion characteristics on the phase noise behavior is analyzed. It could be shown and explained how the phase noise at different offset frequencies to the carrier is affected differently depending on the dispersion characteristic and also the comb width. Design rules for the required comb width as well as the acceptable variation of the produced dispersion from the ideal value depending on the demands for the phase noise improvements could be developed and will be presented.
\end{abstract}


\section{Introduction}
The performance of many applications such as wireless communication or radar is severly limited by the phase noise of their carriers \cite{Coo2012}. With increasing frequencies, the phase noise characteristics that are provided by state-of-the-art systems deteriorate. The phase noise of RF signals generated by classical upconversion schemes follows \cite{Grm2010}
\begin{equation}
\mathcal{L}(f)=\mathcal{L}_0(f)+20\log_{10} m\;,
\end{equation}
where $\mathcal{L}_0(f)$ is the phase noise of the reference signal and $m$ is the upconversion factor. One obvious explaination takes into account the time-domain equivalent of phase noise -- jitter. For ideally upconverted signals, the jitter remains constant but the period of the signal decreases. Hence, the phase noise (i.e. ratio of jitter to signal period) increases. No major improvements of the phase noise of oscillators at lower frequencies could be seen in the past years, e.g. less than 2~dB between \cite{Nic2010} in 2010 and \cite{Ham2019} in 2019.

In order to provide carriers at high frequencies with low phase noise, methods involving optical domain have been developed. Among these, comb lasers and upconversion with electro-optical modulators are most promising \cite{Cha2013}, \cite{Yu2005}. However, this does not have any positive effect on the phase noise. In contrast, the temporal Talbot effect can be used to lower the phase noise during upconversion \cite{Fer2004,Fer2005}. Experiments involving that scheme have been carried out in the past \cite{Jan1981,Neu2010, Zhe2017}. Yet, for practical usability, bulky fiber-based setups are ill-suited. Therefore, an integrated solution should be found. Besides the comb laser, the dispersive element required for the temporal Talbot effect is the key component for this approach and its realization. Analyzing the influence of the dispersive element on the phase noise performance is a key requirement for the component development.

After introducing the frequency upconversion scheme, section~\ref{sec:sim} presents the simulation tool for the phase noise suppression investigations. The following chapter deals with the investigation of different dispersive elements concerning their impact on the phase noise performance. Finally, the results are concluded and an outlook to future work is given.

\section{Principle}
\begin{figure}[htb]
	\centering
	\includegraphics[width=0.9\textwidth]{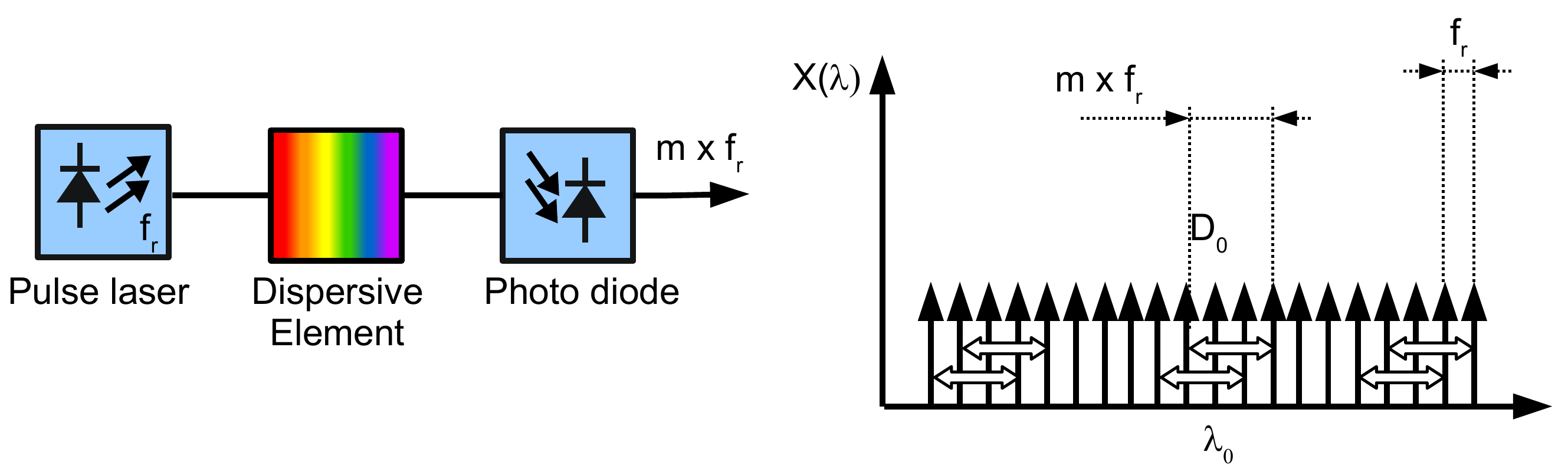}
	\caption{Setup (left) and principle (right) of Talbot effect based upconversion.}
	\label{fig:talbot}
\end{figure}
For the optical upconversion scheme using the temporal Talbot effect, a pulse laser source producing an optical comb spectrum, a dispersive element and a photodiode for opto-electrical conversion are needed (Figure~\ref{fig:talbot}). The spectral lines of the optical comb are spaced by the pulse repetition frequency $f_r$. 
This frequency serves as reference frequency for the upconversion. Being phase coupled, the beating of these lines does not add extra phase noise (in contrast to the beating of two independent laser lines).

Dispersion delays the spectral components of an optical comb with respect to each other. The characteristic dispersion at a wavelength $\lambda$
\begin{equation}
D_\mathrm{char}=\frac{c}{\lambda^2 f_r^2}
\end{equation}
leads to a delay of neighboring comb components spaced by the repetition rate of the pulse laser $f_r$ of one period. A photodiode downconverts all these components resulting in constructive superposition. When a dispersion
\begin{equation}
D=\frac{D_\mathrm{char}}{m}
\end{equation}
is used, this situation is true not for neighboring comb components but for every $m^\mathrm{th}$ component resulting in upconversion. The upconversion factor m is limited by the spectral width of the comb (i.e. by the time-domain pulse width of the pulse laser) \cite{Cha2013}. Using femtosecond lasers \cite{femtoferb}, upconversion factors of more than 1000 are possible \cite{Mel2005}. Being based on interference, no power is lost during upconversion in theory (if the dispersive element does not introduce loss) \cite{Neu2011,Mar2017}.

\begin{figure}[htb]
	\centering
	\includegraphics[width=0.6\textwidth]{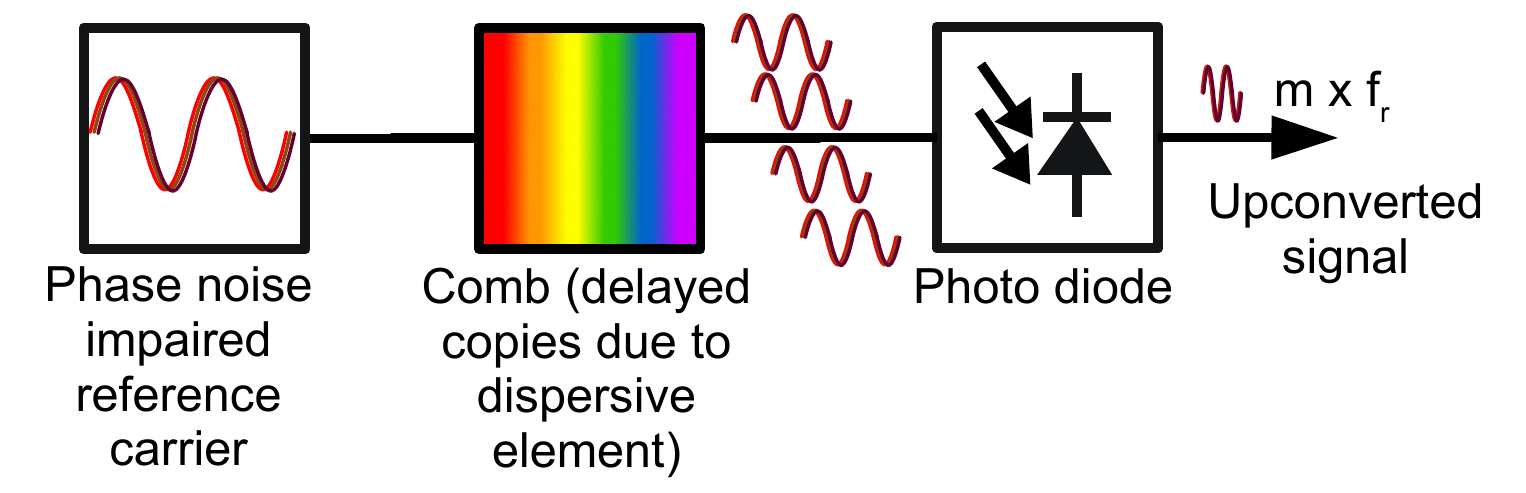}
	\caption{Principle of noise averaging by Talbot effect.}
	\label{fig:pnsim}
\end{figure}

The phase noise of the RF tone produced by the photodiode depends on the phase noise of the repetition frequency of the comb laser $f_r$. Due to the fact that noise (such as phase noise) is decorrelated in time, superimposing two delayed versions of this time domain signal $f_r(t)$ and $f_r(t+T_r)$ with the nominal period of the repetition frequency $T_r=1/f_r$ leads to an averaging that reduces the noise  (Figure~\ref{fig:pnsim}). In broad combs, many lines (e.g. 30000 for a comb width of 3~THz and  a repetition frequency of 100~MHz) are available that are supposed to allow a significant improvement of its noise features.

Phase noise of the comb components can be regarded as a frequency dependent signal $\mathcal{L}(f)$ modulated on the carrier $f_r$. High-frequency noise components have shorter periods and are therefore averaged better within the same time window. In order to average phase noise components near the carrier (i.e. with low frequencies), longer delays are required, e.g. 10~ms for a 10-fold averaging of a component with 1~kHz offset to the carrier. Clearly, the dispersion characteristic will have an effect on the superposition and, consequently, the averaging of the phase noise.

\section{Simulation tool}
\label{sec:sim}
The simulation of phase noise under the presence of chromatic dispersion in a Talbot effect based upconversion scheme is challenging because of the computational and memory requirements. On the one hand, due to the comb, a huge frequency range has to be covered, typically more than 1~THz \cite{femtoferb}. On the other hand, the frequency resolution should be at least around 100~Hz in order to be able to characterize noise close to the carrier. In time-domain, these requirements translate to simulating long sequences with high sample rate. As a consequence, tremendous amounts of memory would be required and the computational effort would be enormous: A 3~THz comb with 100~Hz resolution represented in double (64 bit) values would require more than 446~Gbyte of data -- too much to be stored and processed, especially when different scenarios shall be investigated or parameter sweeps shall be carried out.

However, taking into account the specific properties of the Talbot effect based upconversion will significantly reduce the memory and computing effort:
\begin{itemize}
	\item Most of the data in the spectrum between the comb lines may be neglected because it will not contribute to the phase noise.
	\item All comb lines are approximately identical because of the coherence of the comb source.
\end{itemize}
Thus, the phase noise can be represented in time domain as a noise-impaired signal with the repetition frequency of the comb $f_r$. The degree of oversampling will have an influence on the phase noise that can be modeled and will be investigated in a later step. Also, the time window  $t_\mathrm{sig}$ (length of the signal) is important and determines the frequency resolution, i.e. minimum offset to the carrier that can be investigated. The total time of the signal (time window) determines the frequency resolution (independently of the sampling rate, known as Rayleigh frequency)
\begin{equation}
\Delta f=\frac{1}{t_\mathrm{sig}}
\end{equation}
These kind of signals consume much less memory and can be calculated very fast. For example, a signal with a comb spacing of $f_r=100\;\mathrm{MHz}$ and a 100~Hz resolution represented in double values would just need a bit more than 15~Mbyte. Chromatic dispersion will delay the comb components to each other. That means, the resulting signal after photodetection may be computed by adding delayed copies of the original signal. For a 3~THz comb with 100~MHz spacing, this means 30000 additions which is no problem even for desktop computers. The phase noise $\mathcal{L}(f)$ is calculated from the frequency domain representation of the resulting signal.

\begin{figure}[htb]
	\centering
	\includegraphics[width=0.49\textwidth]{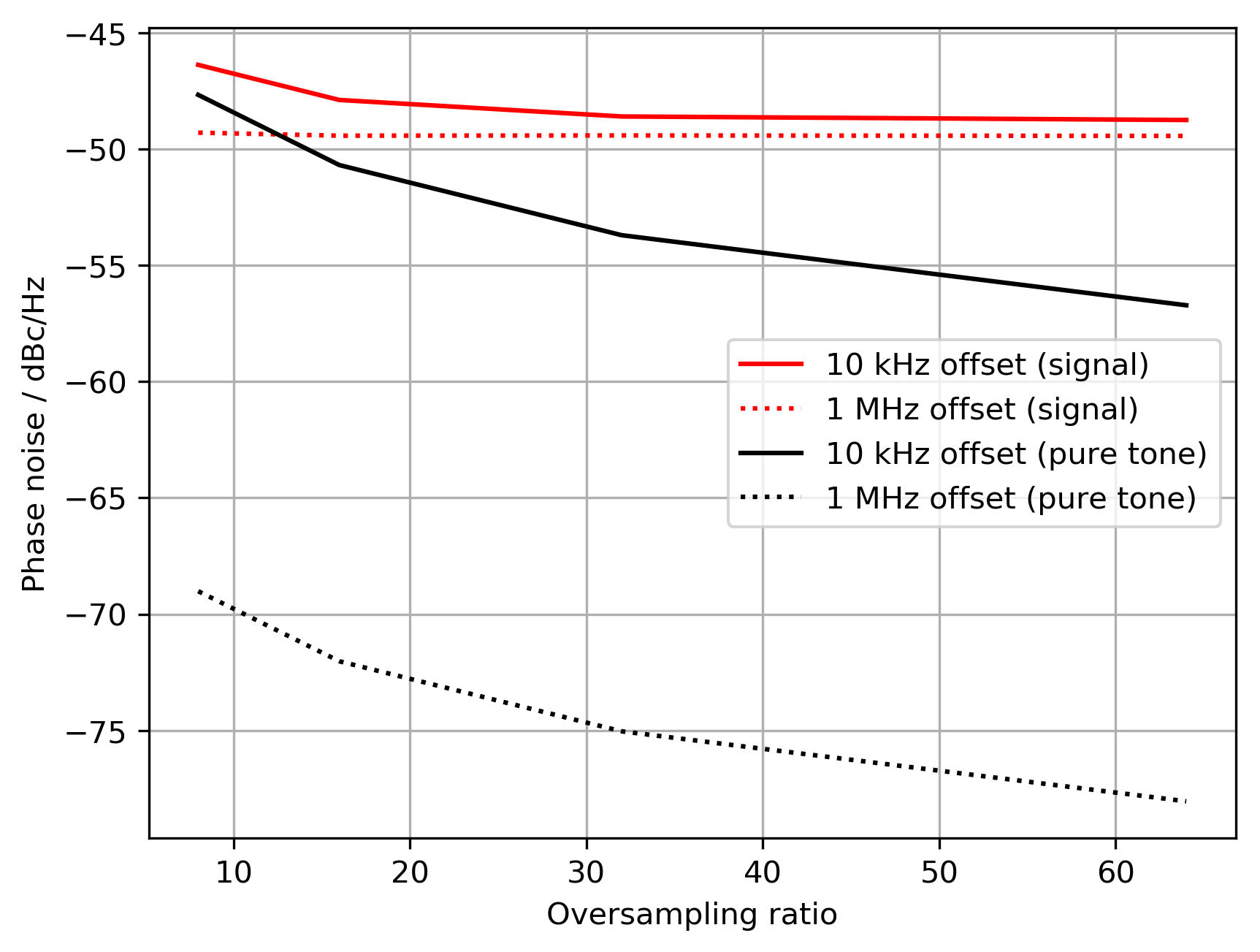}
	\caption{Phase noise at 10~kHz (solid line) and 1~MHz offset (dashed line) to the carrier depending on oversampling ratio for pure tone (black) and phase noise impaired signal (red).}
	\label{fig:simtool_oversampling}
\end{figure}

\begin{figure}[htb]
	\includegraphics[width=0.49\textwidth]{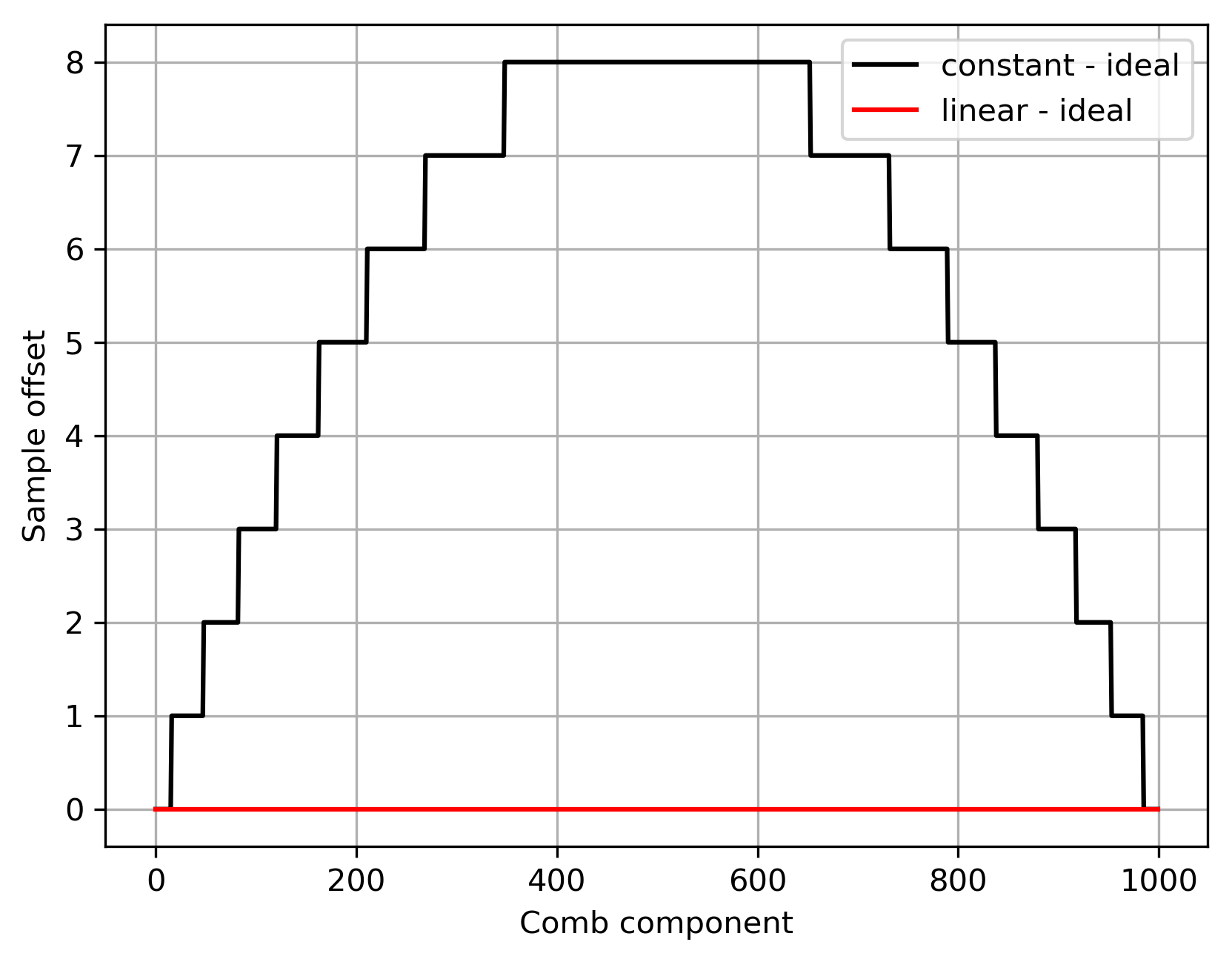}
	\includegraphics[width=0.49\textwidth]{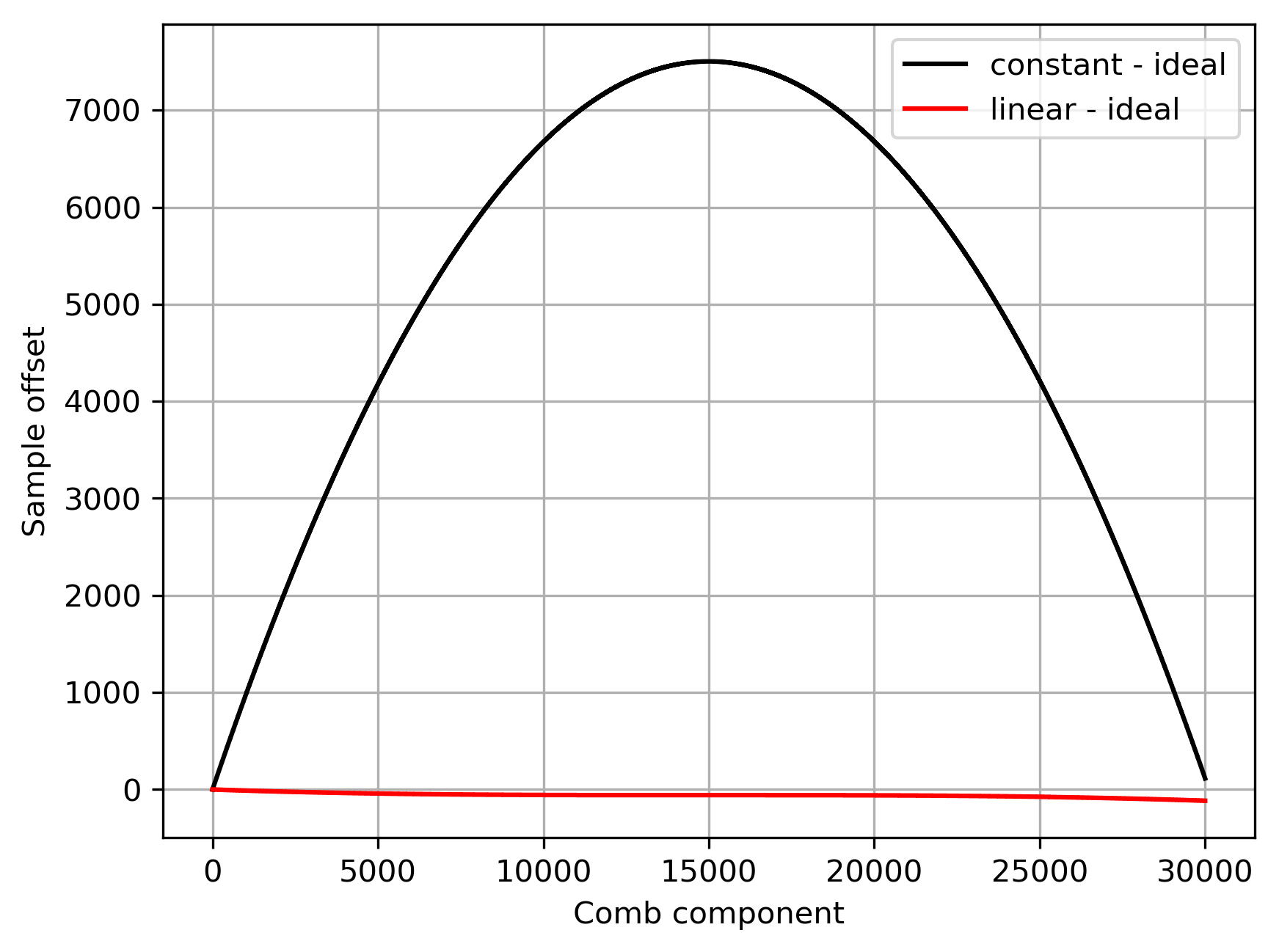}
	\caption{Difference of sample offset between ideal dispersion and linear (red) resp. constant (black) dispersion for a comb width of 100~GHz (left) and 3~THz (right) ($f_r=100\;\mathrm{MHz}$ repetition rate, oversampling factor $N=64$).}
	\label{fig:sampleoffset}
\end{figure}

The described scheme has been implemented. Test sequences with pure tones and carriers with added phase noise have been created. First, the influence of the oversampling ratio has been investigated in order to determine a good compromise between low computational and memory effort and high accuracy. In Figure~\ref{fig:simtool_oversampling}, the phase noise at 10~kHz (solid line) and 1~MHz offset (dashed line) to the carrier depending on oversampling ratio for pure tone (black) and phase noise impaired signal (red) is shown. As expected, the accurracy of the phase noise simulation increases with increasing oversampling ratio for the pure tone. This is due to the higher time resolution equivalent to more samples per period. However, for the phase noise impaired signal it can be seen that for an oversampling ratio above 32, the determined phase noise saturated also for an offset frequency to the carrier of 10~kHz. The phase noise of the pure tone (representing the minimum phase noise that can be simulated) has a significant offset to the phase noise of the signal. That means, the added phase noise dominates and an improvement would be seen.

When the spectral distribution of the phase noise $\mathcal{L}(f)$ is not of interest, its time-domain equivalent jitter may be analyzed
\begin{equation}
J=\int\limits_0^\infty \mathcal{L}(f) \mathrm{d}f
\end{equation}

For the modeling of the delay due to the dispersion, only an integer number of samples will be used. In order to have an offset of one sample between neighboring comb lines, a dispersion
\begin{equation}
D_\mathrm{c}=\frac{\frac{1}{N f_r}}{\frac{c}{\frac{c}{\lambda_0}-f_r}-\lambda_0}
\end{equation}
depending on the oversampling ratio $N$ and the comb repetition frequency $f_r$ at the center wavelength of the comb $\lambda_0$ is required. For a comb repetition frequency $f_r=100\;\mathrm{MHz}$, an oversampling factor $N=64$ and a center wavelength of the comb $\lambda_0=1550\;\mathrm{nm}$ the resulting dispersion is $D_\mathrm{c}\approx 195000\;\mathrm{ps/nm}$. However, the minimum amount of dispersion that will have an effect in the simulation is much lower: It is determined by the delay that causes one sample offset over the spectral width of the comb (i.e. between its spectral lines with the lowest and highest wavelength). For a comb width of $w_\mathrm{comb}=3\;\mathrm{THz}$, this minimum amount of dispersion is 6.5~ps/nm.

\begin{figure}[htb]
	\centering
	\includegraphics[width=0.49\textwidth]{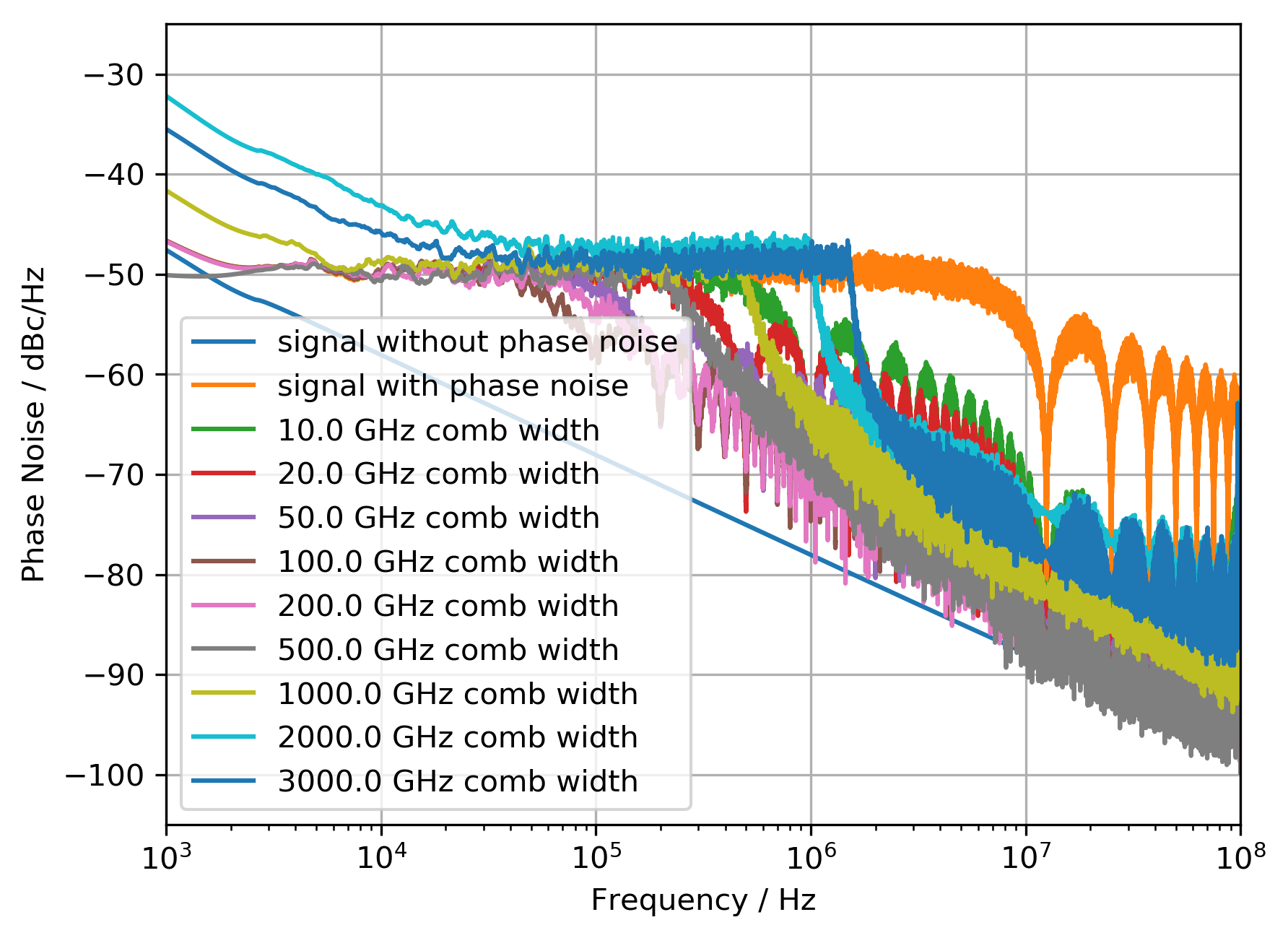}
	\includegraphics[width=0.49\textwidth]{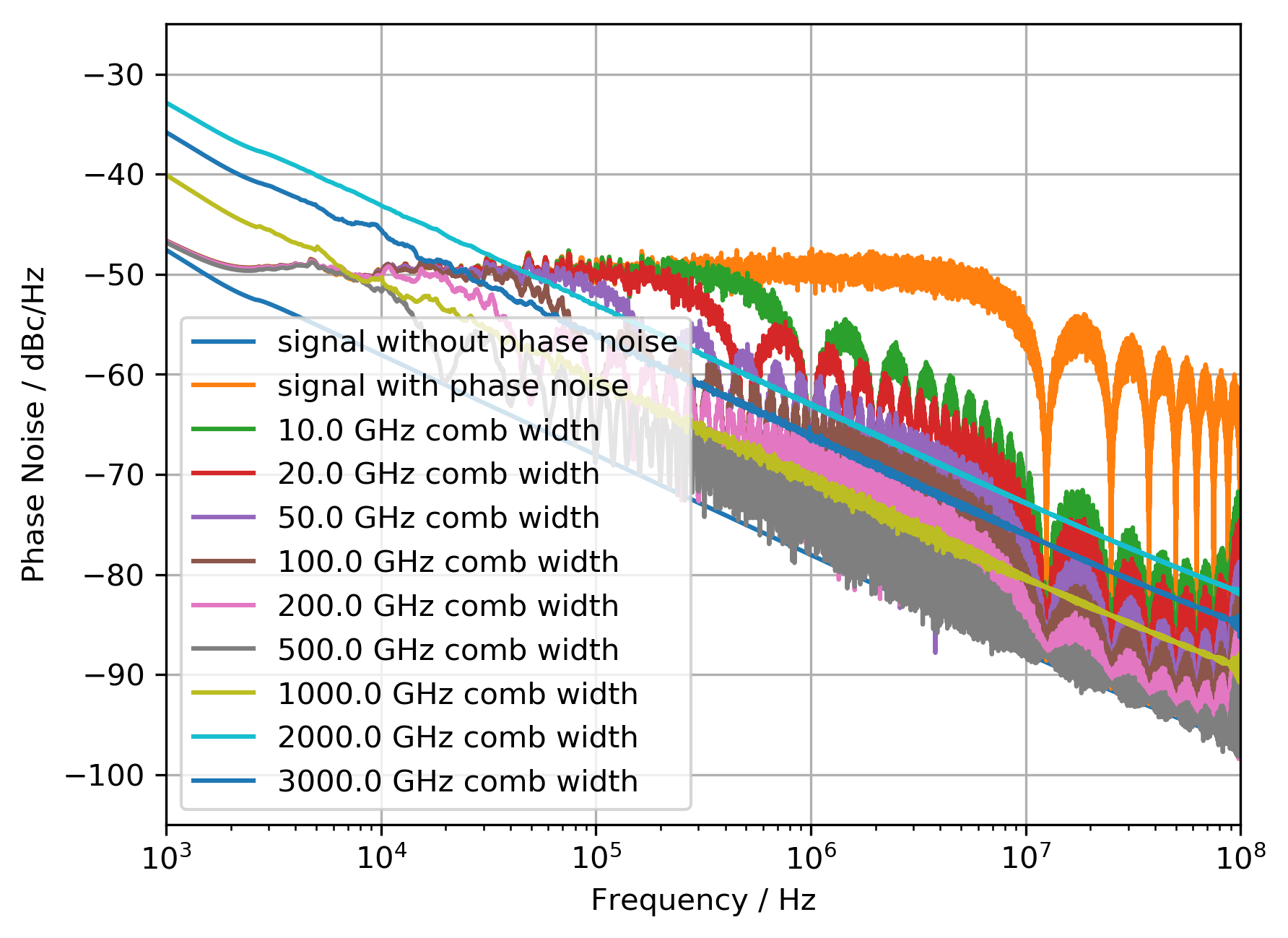}
	\includegraphics[width=0.49\textwidth]{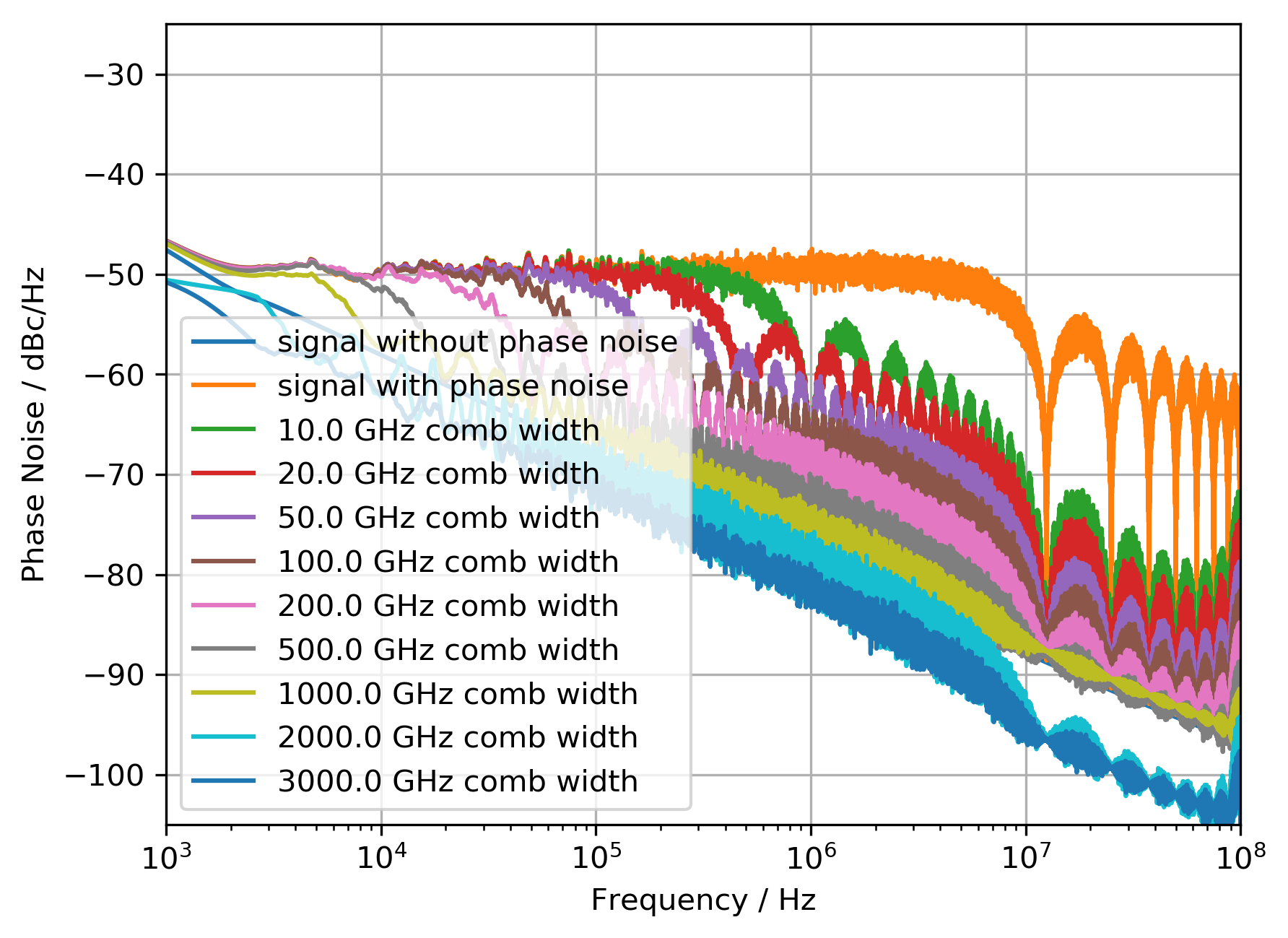}
	\caption{Simulated phase noise with constant (top left), linear (top right) and ideal (bottom) dispersion for different comb widths ($f_r=100\;\mathrm{MHz}$ repetition rate, oversampling factor $N=64$).}
	\label{fig:pn}
\end{figure}
\begin{figure}[htb]
	\includegraphics[width=0.49\textwidth]{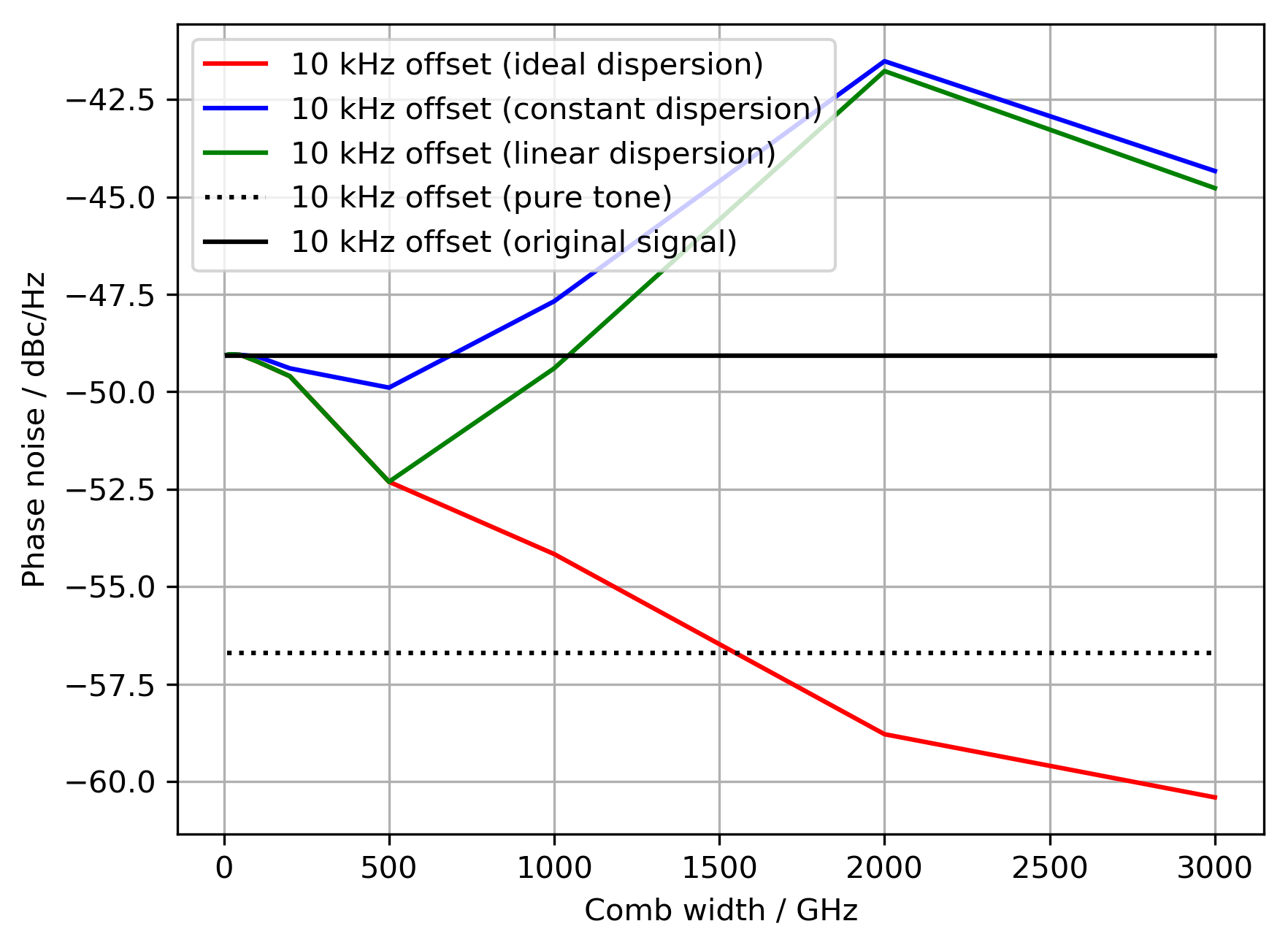}
	\includegraphics[width=0.49\textwidth]{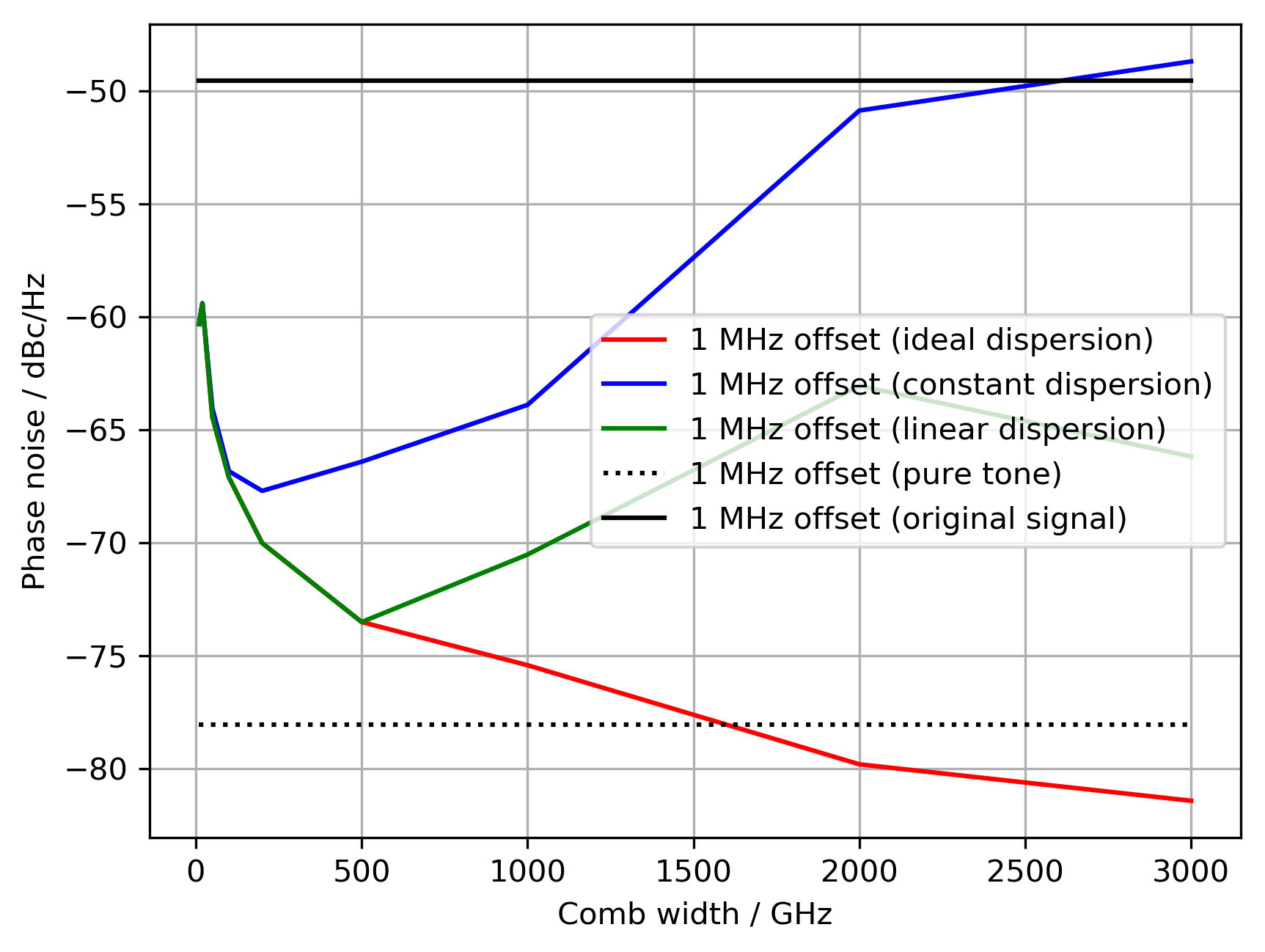}
	\caption{Simulated phase noise at 10~kHz (left) and 1~MHz (right) offset  to the carrier depending on comb width for constant (blue), linear (green) and ideal (red) dispersion ($f_r=100\;\mathrm{MHz}$ repetition rate, oversampling factor $N=64$).}
	\label{fig:pn_10k}
\end{figure}

\section{Dispersion influence}
Three types of dispersion have been investigated that are reasonable for possible implementation: The "ideal dispersion" realizes the optimal dispersion value for each wavelength $\lambda$ in the spectral range of the comb. In order to just investigate the phase noise influence, the upconversion factor of the required dispersion for the Talbot effect is set to $m=1$
\begin{equation}
D_\mathrm{ideal}(\lambda)=D_\mathrm{char}(\lambda)=\frac{c}{\lambda^2 f_r^2}
\end{equation}
The linear dispersion characteristic can be calculated by a first order Taylor series approximation of $D_\mathrm{ideal}$ around the comb center wavelength $\lambda_0$
\begin{equation}
D_\mathrm{lin}(\lambda)=-2\frac{c}{\lambda_0^3 f_r^2} \lambda + 3 \frac{c}{\lambda_0^2 f_r^2}
\end{equation}
Finally, the constant dispersion just realizes the value of $D_\mathrm{ideal}(\lambda_0)$ over the whole spectral range of the comb. Figure~\ref{fig:sampleoffset} compares the difference between the sample offset caused by ideal and linear dispersion (red) as well as the difference between the sample offset caused by ideal and constant dispersion (black) for each comb component. For narrow combs (e.g. 100~GHz in Figure~\ref{fig:sampleoffset}, left) the linear dispersion characteristic is identical to the ideal one within the precision of a full sample offset. Therefore, the same behavior for the phase noise is expected. The constant dispersion characteristic, however, already leads to deviations that should prevent the phase noise reduction from being effective. As the value of the ideal dispersion at the comb center is taken for the constant dispersion, the sample offsets are symmetrical to the center. The mean dispersion of the ideal dispersion characteristic equals the constant dispersion. Therefore, the difference of the sample offsets between ideal and constant characteristic reduce to zero at the minimum and maximum wavelength of the comb. For broader combs (Figure~\ref{fig:sampleoffset}, right), the offsets increase and also the linear dispersion characteristic differs enough from the ideal one to show a significant difference.

The simulated phase noise for the three dispersion characteristics (Figure~\ref{fig:pn}) reveals several remarkable observations:
\begin{itemize}
	\item For smaller combs, only the spectral components of the phase noise farer away from the carrier improves. This is because in narrower combs the maximum time offset between the first and last component is not big enough to enable averaging on a time scale required for low frequencies (e.g. 0.1~ms for 10~kHz).
	\item Constant dispersion and linear dispersion characteristics may not use the full comb width for averaging because the dispersion is not matched for the whole comb. Therefore, phase noise cannot be improved close to the carrier. Additionally, the comb components that are not dispersion matched add noise. That's why with increasing comb width, the phase noise increases, starting from small to large offset frequencies to the carrier.
	\item With ideal dispersion, at a comb width of around 1.5~THz, the noise level of the signal without phase noise is reached. It should be noted that a remaining variation of the numerically retrieved values can be seen in the graphs.
\end{itemize}
In Figure~\ref{fig:pn_10k}, the simulated phase noise at 10~kHz and 1~MHz offset to the carrier is depicted with respect to the comb width depending on the dispersion characteristic summarizing the previous statements.

\section{Conclusion}
With a simulation tool developed to estimate the phase noise influence of dispersive elements on comb laser signals, the effect of different dispersive elements could be shown. For broad combs, e.g. a typical comb laser with a repetition rate of $f_r=100\;\mathrm{MHz}$ and a comb width of 3~THz, precisely realizing the required dispersion is essential. A linear dispersion characteristic will be sufficient for comb widths of less than around 500~GHz while constant dispersion will not be suitable. Phase noise components with larger offset to the carrier are much easier to suppress than components close to the carrier. However, the latter case is the more important one for typical applications. Hence, broad combs and carefully designed dispersive elements are required.

Future work will include the design of integrated dispersive elements based on waveguides, gratings or resonant structures. Using the simulation tool introduced here, it can be investigated how the realized dispersion will affect the phase noise suppression feature. Depending on the comb width, also dispersion characteristics linear with wavelength (that might be easier to realize) are an option.

\bibliographystyle{unsrt}
\bibliography{pn2020_neumann}

\end{document}